# Exceeding the Classical Time-bandwidth Product in Nonlinear Time-invariant Systems

Alireza Mojahed[1*†], Kosmas L. Tsakmakidis[2*], Lawrence A. Bergman[3], Alexander F. Vakakis[1]

[1]Department of Mechanical Science and Engineering, University of Illinois, Urbana, IL 61801

[2]Section of Condensed Matter Physics, Department of Physics, National and Kapodistrian University of Athens, Panepistimioupolis, GR-157 84 Athens, Greece

[3]Department of Aerospace Engineering, University of Illinois, Urbana, IL 61801

*These authors contributed equally in preparing this article
† *Present address*: Massachusetts Institute of Technology, Cambridge, MA 02139, USA

**Abstract**

The classical "time-bandwidth" limit for linear time-invariant (LTI) devices in physics and engineering asserts that it is impossible to store broadband propagating waves (large $\Delta\omega$'s) for long times (large $\Delta t$'s). For standing (non-propagating) waves, i.e., vibrations, in particular, this limit takes on a simple form, $\Delta t\, \Delta\omega = 1$, where $\Delta\omega$ is the bandwidth over which localization (energy storage) occurs and $\Delta t$ is the storage time. This is related to a well-known result in dynamics, namely that one can achieve a high Q-factor (narrowband resonance) for low damping, or small Q-factor (broadband resonance) for high damping, but not simultaneously both. It thus remains a fundamental challenge in classical wave physics and vibration engineering to try to find ways to overcome this limit, not least because that would allow for storing *broadband* waves for *long* times, or achieve broadband resonance for low damping. Recent theoretical studies have suggested that such a feat might be possible in LTI terminated unidirectional waveguides or LTI topological "rainbow trapping" devices, although an experimental confirmation of either concept is still lacking. In this work, we consider a nonlinear but time-invariant mechanical system and demonstrate experimentally that its time-bandwidth product can exceed the classical time-bandwidth limit, thus achieving values, both, above and below unity, in an energy-tunable way. Our proposed structure consists of a single degree-of-freedom nonlinear oscillator, rigidly coupled to a nondispersive waveguide. Upon developing the full theoretical framework for this class of nonlinear systems, we show how one may control the nonlinear flow of energy in the frequency domain, thereby managing to disproportionally decrease (increase) $\Delta t$, the storage time in the resonator, as compared with an increase (decrease) of the system's bandwidth $\Delta\omega$. Our results pave the way to conceiving and harnessing hitherto unattainable broadband and simultaneously low-loss wave-storage devices, both linear and nonlinear, for a host of key applications in wave physics and engineering.

## 1. Introduction and preliminaries

Take any linear time-invariant (LTI) resonant device storing (localizing) for a certain amount of time a wave of amplitude $x$ inside it – this can range [1] from a microcavity in silicon photonics, a meta-atom (e.g., split-ring resonator) in metamaterials, or a metallic/dielectric particle in (nano) plasmonics, to a resonator in a mechanical/acoustic system or in cavity quantum electrodynamics/opto-mechanics. Such systems are nowadays being used in an extremely broad range of applications [1-15] in physics and engineering, from ultrasensitive sensors to on-chip frequency combs, and from ultrafast nanolasers to resonant "meta-atoms" in metamaterials deployed for subwavelength imaging, high-speed modulation, invisibility cloaking, and light slowing/stopping. Moreover, the use of resonators is universal across scales in diverse engineering fields.

Inside this resonant system, a standing wave (or vibration) will oscillate sinusoidally, say with a frequency $\omega_0$, and will decay with time owing to some loss mechanism(s) with a (total) decay rate $\lambda$, i.e., it will satisfy [1] $x(t) \propto \cos(\omega_0 t) \times e^{-(1/2)\lambda t}$. Hence, in the resonance approximation and in the usual underdamped regime ($\lambda/2 \ll \omega_0$), the intensity of the vibration will be given by,

$$E(\omega) = \frac{1}{(\lambda/2)^2 + \omega^2}, \tag{1}$$

where in the frequency domain the energy spectral density is defined as $E(\omega) = |\mathcal{F}\{\langle \dot{x} \rangle\}|^2$ with $\langle . \rangle$ indicating the operator extracting the envelope of its argument and $\mathcal{F}$ denotes the Fourier-transform operator. From relation (1) it is immediately seen that the (classical half-amplitude) bandwidth is $\Delta\omega = \lambda$; in other words, the product of the storage time, $\Delta t = 1/\lambda$, with $\Delta\omega$ appears to always be equal to unity, $\Delta t \Delta\omega = 1$, and this holds for any single-degree-of-freedom lightly damped linear time-invariant resonator. This limit is as the classical "time-bandwidth (T-B) limit" [1,9,11,16,17] of LTI resonant devices.

It is therefore intriguing to inquire what happens when only the 'L' (linearity) assumption is broken, i.e., when a system becomes nonlinear but still time-invariant. Right from the start, however, we immediately recognize that in the nonlinear case the notion of bandwidth is not readily available, so in order to study the time-bandwidth product of such a nonlinear time-invariant (NTI) resonator we should first appropriately define its "nonlinear bandwidth." We show in the following that for a general class of dynamical nonlinear (e.g., communication [18]) systems this task can be achieved by considering the "root mean square (RMS) bandwidth" of the system. A detailed discussion of the nonlinear bandwidth definition is left for another work, and here we only provide a brief exposition upon which our later results can be based.

To this end, we note that extending the concept of bandwidth for general classes of dynamical systems requires a formulation that is valid not only for low-loss LTI systems (for which the classical definition applies) but also encompasses systems with more general features, e.g., moderate or even high dissipative rates, time variant properties and nonlinearities. Hence, we begin by acknowledging the fact that the bandwidth of a dynamical system (as in the traditional classical half-amplitude bandwidth definition) is one of its inherent properties and relates to its energy ($E(\omega)$) dissipation capacity (loss rate). For instance, in the case of a SDOF LTI resonator, a larger bandwidth, i.e., larger energy dispersion in the frequency domain, translates to faster energy dissipation, hence to higher time-locality of the energy signal. This means that, e.g., in an acoustic setting, a larger bandwidth resonator connected to an acoustic waveguide releases its stored energy in the form of time localized (broadband) waves at a faster rate to the waveguide; conversely, a smaller bandwidth resonator releases its stored energy to the waveguide in the form of narrowband waves at a slower rate. The time-locality of a signal can be quantified by its temporal variance, $\sigma_t^2$, while its frequency dispersion can be quantified by its frequency-domain variance, $\sigma_\omega^2$. As it is well established, these two quantities are related through the Fourier uncertainty principle [19], which states that for a signal (satisfying the Fourier-transformation requirements) the value of $\sigma_t^2 \sigma_\omega^2 = C \geq 1/16\pi^2$, where $C$ is a constant. With the inverse proportionality of $\sigma_t^2$ and $\sigma_\omega^2$ established, we now note that $\sigma_\omega^2$ is, e.g., in standard communication systems [18], also known as $(\Delta\omega_{rms})^2$ – the aforementioned RMS bandwidth [2]. With this (RMS) bandwidth formula and its inverse relation to the time-locality of the signal, we can now compute the bandwidth of broader classes of dynamical systems, such as, both, single- and multi-degree-of-freedom linear, time variant or invariant, as well as nonlinear dynamical systems. Indeed, one more argument that this general RMS bandwidth definition is the appropriate one, is that for the energy spectrum $E(\omega)$ of the signal in Eq. (1), the RMS bandwidth given by,

$$\Delta\omega_{rms} \equiv \Delta\omega = 2\sqrt{\frac{\int_{-\infty}^{\infty} \omega^2 E^2(\omega) d\omega}{\int_{-\infty}^{\infty} E^2(\omega) d\omega}}, \qquad (2)$$

leads, by a simple substitution of Eq. (1) to Eq. (2), to the correct result for a linear SDOF system, that is, $\Delta\omega^* = \lambda$ – as it should, where the superscript (*) is used to denote the classical bandwidth definition. We emphasize here that since $\Delta\omega$ in (2) quantifies the overall *energy* dissipation rate of a system, the argument on the right-hand side contains $E^2(\omega)$ – and not other powers of $E(\omega)$ since these do not represent any physical feature(s) of the system.

The classical T-B limit ($\Delta t \Delta\omega = 1$) has never been violated in any LTI resonant or wave-localizing structure [1,9,11], including (but not limited to) in such broad areas of research as optical microcavities, metamaterial or dielectric resonant structures, (nano)plasmonics, and Anderson localization of waves – remaining a long-sought-after fundamental objective in wave physics and engineering to do so. Achieving T-B products above unity would, e.g., imply that we could harness the "best of both worlds" in the above systems, namely storing waves for, simultaneously, long times *and* broad bandwidths. Conversely, tuning this limit *below* unity would mean that, e.g., for a fixed large bandwidth, the release of a wave excitation would be faster than what is normally allowed (for that bandwidth); again, this would find a range of important technological applications, such as energy harvesting [3], vibration absorption and isolation [12], or the attainment of on-chip nonreciprocity [7,13].

Based on the previous nonlinear bandwidth conception we will demonstrate experimentally that the T-B product of a NTI mechanical system can be *tuned* – both *above* and, even more intriguingly, *below* unity – as desired. In our concept (cf. Fig. 1a), a nondispersive acoustic waveguide is rigidly coupled to a SDOF nonlinear oscillator of mass $I$, stiffness coefficient $\omega_0^2 I$, viscous damping coefficient $\lambda I$, dry friction coefficient $\mu I$, and linear natural frequency $\omega_0$. The sources of nonlinearity in this system are Coulomb friction (weak nonlinearity), inelastic impacts [12] (strong nonlinearity) of the mass at a rigid barrier (stop) situated at clearance $\delta$, and repulsive magnetic forces when additional magnets are attached to it and to the rigid stop. Depending on the occurrence (at higher energies) or absence (at lower energies) of impacts, the oscillator response is denoted by $x_{nl}(t)$ or $x_l(t)$, respectively. In this system, it may readily be proved that the difference between the in-coupled waves (i.e., incident waves from the waveguide to the resonator) and out-coupled waves (i.e., outgoing waves from the resonator to the waveguide) is only affected by the response of the oscillator [20]; hence, it suffices to examine only the dynamics and time-bandwidth characteristics of the nonlinear oscillator itself. Such a system enables us to disproportionally decrease (increase) its decay-time constant (storage time), $\Delta t$, compared to a corresponding increase (decrease) of its bandwidth $\Delta \omega$ (cf. Figs. 1b-d) to achieve a time-bandwidth product ($\Delta t \Delta \omega$) with values either above or below unity ($\Delta t \Delta \omega \lessgtr 1$).

## 2. Results

As an experimental realization of the SDOF oscillator of Fig. 1a we have analytically studied, computationally modeled, and experimentally characterized a NTI resonator capable of possessing a time-bandwidth product either *below* or *above* unity. Referring to Fig. 2, the experimental system consists of a pendulum of length $L$ (which is relatively long to achieve slow measured dynamics, i.e., at low frequencies), mass $M$, moment of inertia $I$, and natural frequency $\omega_0$. While at rest, the pendulum mass is situated away from a rigid barrier (stop) at distance $\delta$. Depending on the initial amplitude (energy) of the pendulum, impacts may or may not occur between the pendulum mass and the rigid stop. In this configuration the sources of energy dissipation are Coulomb friction and viscous damping with coefficients $I\mu$ and $I\lambda$, respectively, originating from the pivot of the pendulum), and, more importantly, the inelastic impacts of the pendulum at the rigid stop. Accordingly, the angle of the pendulum is related to the translational displacement of the oscillator of the model of Fig. 1a simply by $\theta(t) \approx x(t)/L$. A detailed description of the experimental fixture and its mathematical reduced-order model are discussed in the Appendix. At a later phase of the experiments, a set of magnets was attached both at the site of the rigid stop and on the mass of the pendulum giving rise to repulsive magnetic forces between the pendulum and the stop; this arrangement generated softening nonlinear forces whose effects on the T-B product we wish to explore. In what follows, we explore the system response in three different configurations, namely where (i) no impacts occur while the system oscillates (case 1); (ii) inelastic impacts occur at the rigid barrier (case 2); and (iii) inelastic impacts occur in the magnetic field generated by the magnets (case 3).

Case 1 (no impacts) serves as a baseline to study the effect of energy tunability (nonlinearity) on the time-bandwidth characteristics of such a nonlinear pendulum (oscillator). Initially, the system is set up in such a way that no impacts occur between the oscillator and the barrier, i.e., the considered clearance value $\delta$ is large. Hence, the only source of nonlinearity in this case is the weak friction originating from the joints of the experimental fixture, as mentioned before. The time-domain response of the system and its corresponding wavelet transform are shown in Figs. 3a and 3b, respectively, where we also compare the experimental and corresponding computational (simulation) results of Case 1 (in Fig. 3a) – finding excellent agreement between them. The wavelet transform spectrum in Fig. 3b elucidates the frequency content of the signal. It reveals a fundamental harmonic of constant frequency (~0.5 Hz), and additionally, owing to the weak friction nonlinearity, small-amplitude 3$^{rd}$ and 5$^{th}$ harmonics at ~1.5 Hz and ~2.5 Hz, respectively – both, playing a minor role in the response. Note that, unlike the viscously damped linear oscillator, here, due to the nonlinearity of the system originating from dry (Coulomb) friction, its bandwidth and decay-time constant are energy dependent and tunable. However, as seen from Fig. 4c, even though the system is energy-tunable (nonlinear), its time-bandwidth product does not vary appreciably with respect to the input energy, remaining almost equal to unity throughout – cf. the dash-dotted green curve in Fig. 4c.

For case 2, we set the clearance to $\delta = 3$ cm, so that impacts between the pendulum and the rigid barrier can occur for relatively large input energies. The sharp changes in the velocity signal observed in Fig. 3c, and the broadband bright-colored regions in its corresponding wavelet transform spectrum seen in Fig. 3d, indicate the occurrence of such impacts. In terms of the dynamics of the system, the impacts cause the system to possess "hardening stiffness nonlinearity" [12], that is, it causes the fundamental frequency of oscillation to decrease with decreasing energy – or, equivalently, increasing time (see inset in Fig. 3d). The impacts release local "bursts" of energy (cf. vertical bright-colored bands in Fig. 3d) resulting, together with the inelasticity of the impacts, in the energy of the system decreasing after each impact, and giving rise to a multitude of high-frequency harmonics as may be seen in Fig. 3d. Overall, this intricate phenomenon leads the system to possess much higher dissipative capacity (i.e., higher *bandwidth*, $\Delta\omega$ – cf. red error bars and blue curves in Fig. 4a). Furthermore, the decay-time constant of the system decreases because the hardening nonlinearity increases the frequency of oscillation, thereby increasing the rate of energy dissipated by the damper.

Crucially, however, as illustrated in some detail in the Appendix and Ext. Data Figs. 1 and 2, the decay rate associated with the generated high harmonics is disproportionally large compared with the decay rate characterizing the fundamental harmonic; overall, this results in the total decay rate (storage time, $\Delta t$) to increase (decrease) appreciably more than the aforementioned increase of the bandwidth $\Delta\omega$ of the system – as a result of which the time-bandwidth product, $\Delta t\Delta\omega$, of the system in this case falls *below* unity (see Fig. 4c).

In Case 3, by contrast, we show that by slightly modifying the experimental configuration of Case 2 we may conceive a dynamical system whose time-bandwidth product can now be energy-tuned to possess values both above *and* below unity. Here, compared to Case 2, we add three sets of magnets, namely, two pairs of magnets placed close to the location of the rigid barrier, and an additional magnet directly attached to the mass of the pendulum – cf. Fig. 2, in such a way that the magnets on the barrier repel the magnet on the pendulum. These repulsive magnetic forces introduce a "softening nonlinearity" to the dynamics of the pendulum, whereby its frequency of oscillation increases with decreasing energy, which, in turn, decreases the energy dissipated by the system owing to the reduction of the angular velocity; that is, the decay rate decreases, or the storage time $\Delta t$ *increases* – cf. black error bars in Fig. 4b. Simultaneously, as may be seen from Fig. 4a (black error bars), the bandwidth $\Delta\omega$ of the system decreases because the softening nonlinearity decreases the overall energy dissipation of the oscillator – but, crucially, as shown in Ext. Data Figs. 3 and 4 in the discussion in the Appendix, this decrease in $\Delta\omega$ is not so large compared to the increase of the storage time, $\Delta t$. Hence, we have the exact opposite situation to Case 2 (hardening nonlinearity), so the time-bandwidth product $\Delta t\Delta\omega$ characterizing the system may, in this case, exceed unity. For yet higher input energies, the effect of the magnets becomes practically inconsequential, as a result of which we revert, for this system as well, to Case 2, and the T-B product can now also become less than unity.

All of the above properties can be verified by examining the wavelet transform spectra of the responses of this system (Fig. 3e and Fig. 3f), as well as the dynamics of each harmonic (Ext. Data Fig. 3 and 4 in the Appendix) for different ranges of energy, that is, for different times. From Fig. 3e we discern that at early times, i.e., for high energies, the dynamics is dominated by the hardening nonlinearity (similarly to Case 2) caused by the impacts, as the frequency decreases with time in that regime (Fig. 3f up to ∼ 6 s). However, after ∼ 6 s, as explained above, the dynamics is anticipated to be (and in fact is) dominated by the softening nonlinearity because no impacts occur thereafter, since the only source of nonlinearity affecting the oscillation of the system are the softening magnetic forces. As time lapses further, even the softening nonlinear effects die out, and the oscillation frequency becomes almost constant – in that regime the dynamics is dominated solely by the weak frictional nonlinearity of the system. These effects can also be clearly seen by examining the dynamics of, e.g., the two leading harmonics of the system (Ext. Data Fig. 3 in the Appendix). For instance, for the softening regime of Case 3 (between ∼ 6 s and ∼ 15 s in Fig. 3f), we may see from Ext. Data Fig. 3 that the lifetimes of the harmonics *increase* compared to the same harmonics of Case 2 (Ext. Data Fig. 1), leading to an overall increased storage time $\Delta t$, which, combined with the slight decrease in the system's bandwidth of the system $\Delta \omega$ owing to the nonlinearity, ultimately leads to T-B product above unity, as shown in Fig. 4c (black error bars). Similar conclusions can be drawn for all other points corresponding to Case 3 in Fig. 4c, by examining the temporal evolution of the corresponding harmonics.

## 3. Discussion

The configuration considered in Case 3 allows for more flexibility and tuning (with energy) of the T-B product of the system, both above and below unity. When the input energy is high enough for impacts to occur (i.e., larger than the "critical" energy of $\sim 10^{-2}$ J kg$^{-1}$m$^{-2}$), the T-B product of the system is less than unity, as explained above. However, when the same system is excited with input energy less than the critical value, its T-B product becomes larger than unity, owing to the softening nonlinear effects outlined earlier. Finally, the T-B product approaches unity as the input energy decreases even further and reaches an approximately linear regime (with small perturbations provided by friction at the pivot).

This work demonstrates experimentally and theoretically that the T-B product of NTI systems can attain values both above and below the classical limit of unity. The key element to achieving this result was to judiciously engineer the nonlinear flow of energy in the frequency domain (through the harmonics generated by the nonlinearity) in a tunable (with energy) way for this system. In fact, for a general class of suitable NTI systems and for sufficiently high excitation intensities (which are needed to trigger nonlinear effects), one may exploit hardening nonlinearities to make the lifetime of the harmonics decrease faster than the increase of the bandwidth of the system, leading to the T-B product falling below unity. By contrast, for smaller energies one may exploit softening nonlinearities where the opposite effect occurs, namely, the lifetime of the harmonics increases faster than the decrease of the bandwidth of the system, leading to the T-B product exceeding unity. Hence, our results open the possibility of conceiving resonant time-invariant systems – ubiquitous throughout wave physics and engineering [9,21] – where we could harness, in a tunable way, the "benefits of both worlds". That is, devices that store waves for, simultaneously, long times and broad bandwidths, or resonators with simultaneously broadband resonances with low dissipation rates (or high Q-factors), with a plethora of hitherto unattainable opportunities for stronger wave-matter interactions, and for buffering, processing and harnessing various types of waves and vibrations.

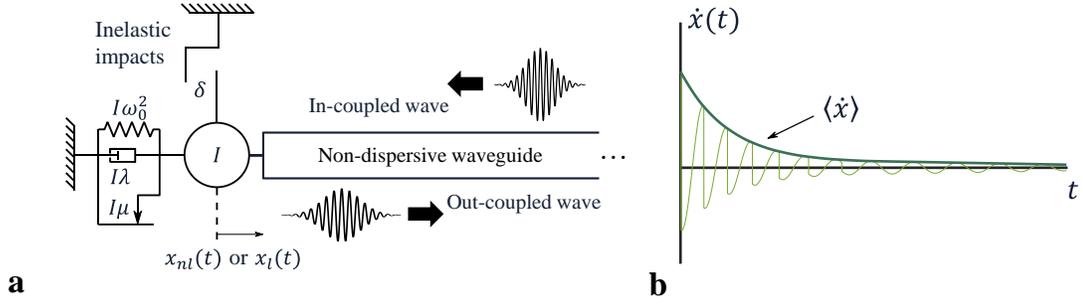

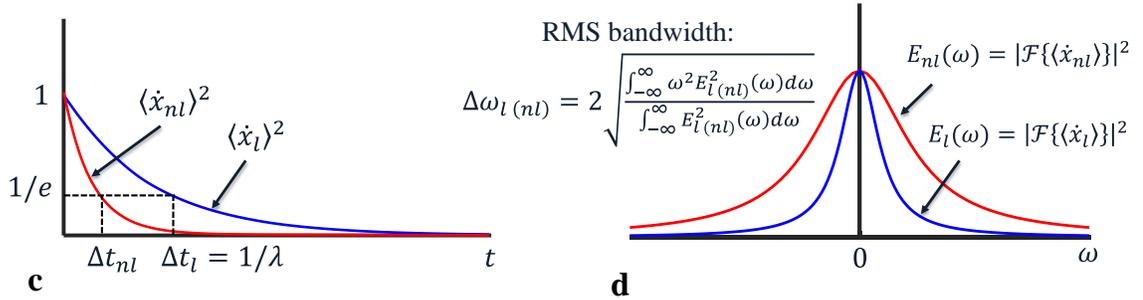

Fig. 1. Scheme for tunable time-bandwidth product in a nonlinear time-invariant system. Basic definitions and the deployed nonlinear time-invariant system are shown: (a) Basic concept of a nondispersive waveguide connected to an oscillator with mass $I$, linear stiffness coefficient $I\omega_0$, viscous and friction coefficients $\lambda I$ and $\mu I$, and linear natural frequency $\omega_0$. Inelastic impacts [12] occur at a rigid barrier (stop) situated at clearance $\delta$ from the oscillator. Depending on the occurrence (at higher energy) or absence (at lower energy) of impacts, the oscillator response is denoted by $x_{nl}(t)$ or $x_l(t)$, respectively. The temporal (or frequency-domain) difference between the in-coupled and out-coupled waves is solely determined by the response of the oscillator [18] (at their connection point). Hence, the response of only the oscillator suffices for studying the time-bandwidth characteristics of this system. (b) Free response (velocity, $\dot{x}$) of the nonlinear oscillator attached to the waveguide (light green curve), and its envelope $\langle \dot{x} \rangle$ (dark green curve). (c) Storage time (decay-time constant) of the square of the envelopes of, both, a linear oscillator (in the limit $\mu = 0, \delta \to \infty$ – blue curve) whose response is denoted by $x_l$, and a nonlinear oscillator with friction and undergoing inelastic impacts (red curve), similar to the one considered in (a) whose response is denoted by $x_{nl}$. Note that the squared values of the velocity envelopes are considered, since they are proportional to the corresponding total energies. (d) Energy spectral densities of $\langle \dot{x}_l \rangle$ and $\langle \dot{x}_{nl} \rangle$, that is, $E_l(\omega)$ (blue curve) and $E_{nl}(\omega)$ (red curve), respectively, and their associated self-consistent root mean square (RMS) bandwidths [2], $\Delta\omega_l$ and $\Delta\omega_{nl}$.

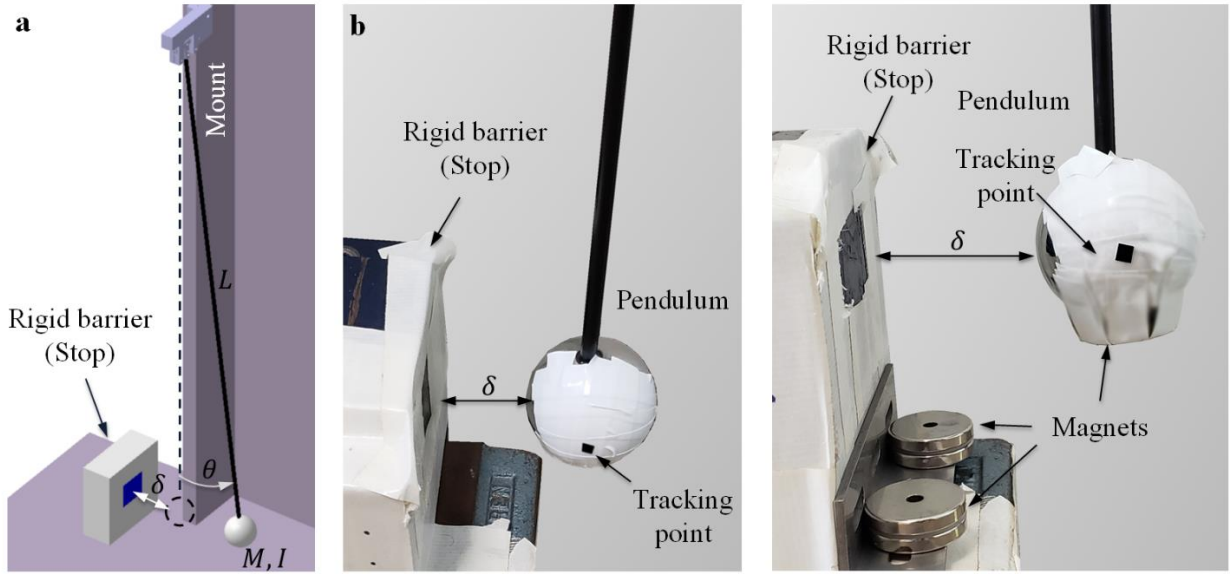

Fig. 2. Experimental configuration. (a) The experimental realization of the oscillator of Fig. 1a consists of a carbon fiber rod of length $L$ (black), a steel ball of mass $M$ attached at the end of the rod (light gray) producing a moment of inertia $I$, a rigid barrier (light gray), and an impact area (dark blue). (b) Experimental fixture for the considered Case 2, where steel-to-steel inelastic impacts occur – the clearance between the barrier and the pendulum at rest is equal to $\delta$. (c) Experimental fixture for Case 3, where two pairs of permanent magnets are positioned close to the rigid barrier and an additional magnet is attached to the pendulum mass, generating repulsive magnetic forces between the barrier and the pendulum. These forces are superimposed to the steel-to-steel inelastic impact forces acting on the pendulum mass.

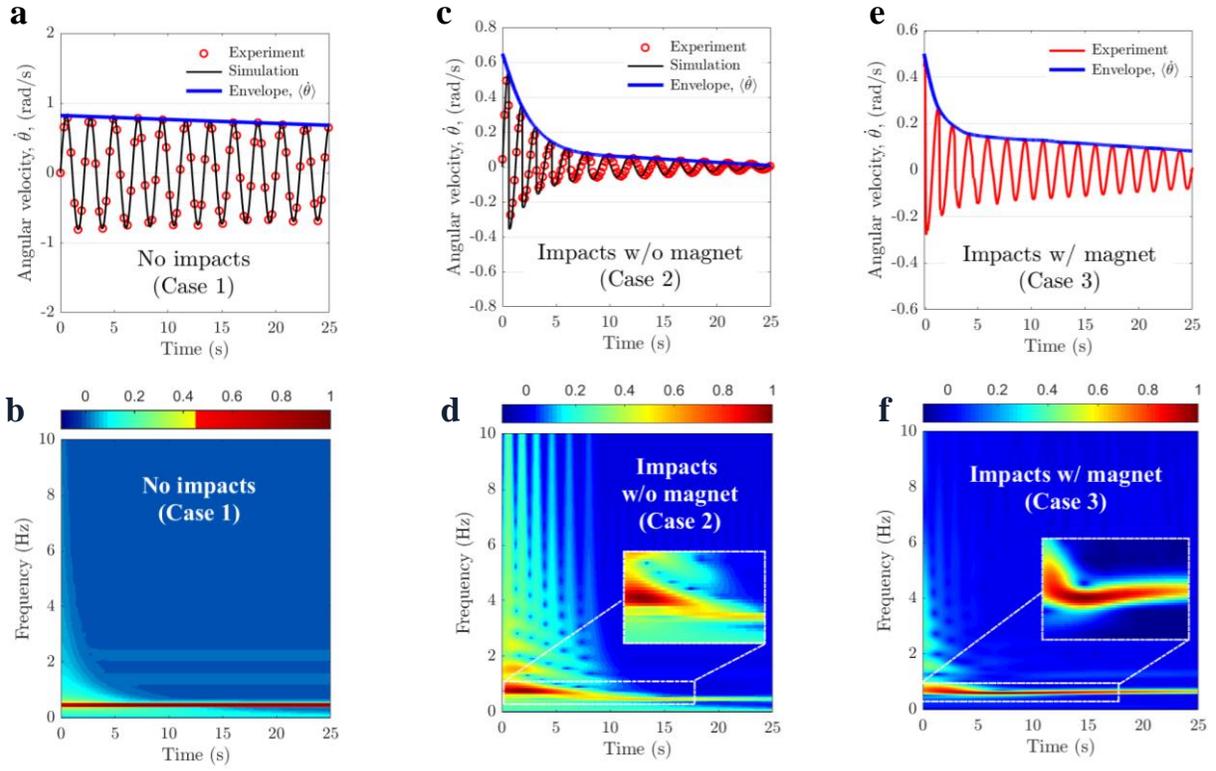

Fig. 3. Angular velocity time series and their corresponding wavelet spectra [7] of the three Cases considered. (a) Simulated angular velocity of the pendulum for Case 1 (no impacts occur) when released from an initial angle of $\theta_0 \sim 15°$ (black curve), along with the corresponding experimental angular velocity of the pendulum (red circles), and their envelope (blue curve). (b) Wavelet transform magnitude (normalized in the range (0,1) by its maximum value), of the experimental response of the pendulum in (a). (c) Angular velocity of the pendulum for Case 2 (impacts occur) when released from an initial angle of $\theta_0 \sim 11°$ (black curve), along with the corresponding experimental angular velocity of the pendulum (red circles), and their envelope (blue curve). (d) Normalized wavelet transform magnitude of the experimental response of the pendulum in (c). (e) Experimental angular velocity of the pendulum for Case 3 (impacts occur in the presence of the magnetic field) when released from an initial angle of $\theta_0 \sim 9°$ (red curve), along with its envelope (blue curve). (f) Normalized wavelet transform magnitude of the experimental response of the pendulum in (e).

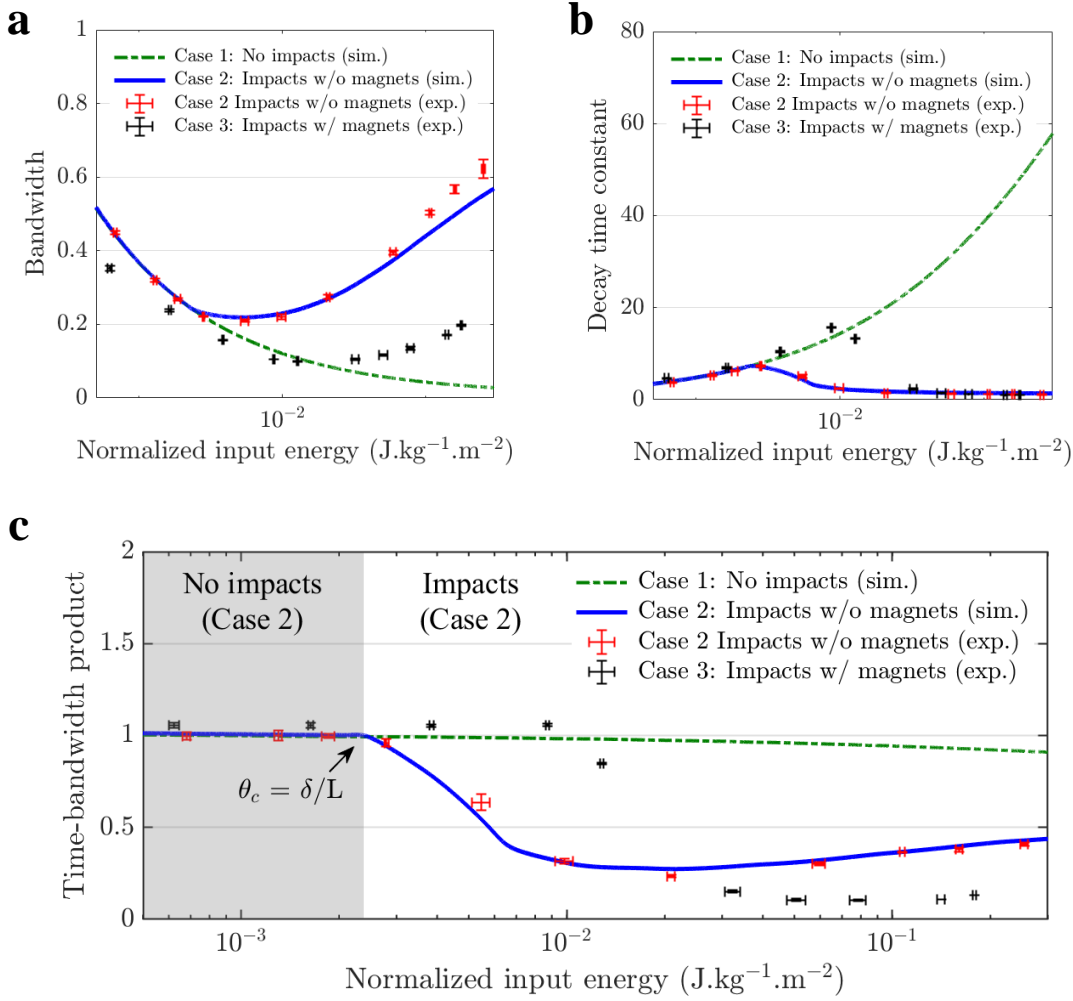

Fig. 4. Time-bandwidth characteristics of the nonlinear time-invariant system. (a) Simulated bandwidth of the pendulum without impacts (Case 1) (green curve), and with impacts (Case 2) (blue curve), validated by the corresponding experiments (red error bars). The dissipative capacity (bandwidth) of the system increases compared to Case 1 once the impacts commence. Black error bars represent the experimentally measured bandwidth for Case 3, showing that the softening nonlinearity introduced by the magnetic force *reduces* the bandwidth of the system compared to Case 2. (b) Simulated decay-time constant (storage time) of the system for Case 1 (green curve), along with the simulated and experimental decay-time constant for Case 2 (blue curve and red error bars, respectively) and the experimental decay-time constant for Case 3 (black error bars). (c) Time-bandwidth product for the cases considered in (a) and (b). The plots of the time-bandwidth product for Case 1 (simulation: green curve) and Case 2 (simulation: blue curve; experiment: red error bars) show that it goes significantly below unity once the nonlinearity engages ($\theta_0 > \theta_c$), whereas before that it is equal to the time-bandwidth product of the system of Case 1 as expected. Black error bars show that by adding the magnetic field, i.e., the softening nonlinearity, the time-bandwidth product can now be pushed both *above* unity for lower energies and, *below* unity for higher energies once the impacts occur; hence, the tunability with energy of the time-bandwidth product can be proven.

**Appendix**

**A1. Bandwidth of tunable-with-energy nonlinear time-invariant (NTI) systems**

The well-known (half-amplitude) definition of bandwidth in a linear time-invariant system considers the frequency response function – i.e., the Fourier transform of the free (impulse) response – of the system. For a low-loss linear system, this quantity is equal to its overall dissipative capacity – in this case, typically the viscous damping coefficient. Likewise, to self-consistently compute the bandwidth of a tunable-with-energy NTI system, we may start by acknowledging that, in effect, the bandwidth of a system indicates how much its energy is localized in time; in turn, this is directly related to how fast (or slow) its energy is dissipated. Hence, instead of considering the Fourier transform of the free response of a nonlinear system (which, strictly is not applicable since the response is non-stationarity), or its frequency response function (which, again, strictly is not applicable due to the generation of higher harmonics in the response), one can compute the well-known root mean squared (RMS) bandwidth [2] of the system using its energy. However, since computing the energy of a nonlinear system is not necessary always possible, e.g., for an experimental system whose parameters are not identified, instead one may consider the envelope of its kinetic energy [22], which is sufficient given that the kinetic energy is proportional to the velocity-squared of the system (factored by the "mass" of the system); clearly, measuring the envelope of the velocity time series is always possible in both simulations and experiments. Owing to the fact that the envelope of the velocity corresponding to the free response of a nonlinear system is monotonically decaying and does not possess any non-stationary frequencies, its Fourier transform can be computed and used to compute the bandwidth of the system as shown in Eq. (2) in the main body of the work.

As discussed in the main text and seen from the results of Fig. 4a, nonlinearity, causing energy tunability, directly affects (increases or decreases) the bandwidth of an NTI system. The effect of nonlinearity reveals itself clearly in the wavelet transform spectrum of the corresponding response, representing the evolutions of the dominant harmonics in time. In our system, this translates to the presence of numerous harmonics of the fundamental frequency, as may be seen by the wavelet transform spectra in Fig. 3. Because of this, for cases 2 and 3 we can express the angular velocity of the pendulum as $\dot{\theta}(t) = \sum_{i=1}^{\infty} \dot{\theta}_i(t)$, where $\dot{\theta}_i(t)$ is the component of the angular velocity due to the $i$-th harmonic. Subsequently, the envelope of the angular velocity of the pendulum can be expressed as,

$$\langle \dot{\theta}(t) \rangle = \sum_{i=1}^{\infty} \alpha_i \langle \dot{\theta}_i(t) \rangle, \tag{3}$$

where $\langle \dot{\theta}_i(t) \rangle$ is the envelope of each harmonic component of the angular velocity, and $\alpha_i$ is associated with the phase difference between the fundamental (1$^{\text{st}}$) and the $i$-th harmonic of the velocity. Substituting Eq. (3) into Eq. (2), we find that the "nonlinear bandwidth" of the system, i.e., the bandwidth of the signal $\dot{\theta}(t)$, is given by,

$$\Delta \omega^{*2} = \frac{\sum_{i=1}^{\infty} \alpha_i^4 E_i^2 \Delta \omega_i^{**2}}{E_0^2} + 4 \frac{\int_{-\infty}^{\infty} \left[ \omega^2 \left( |\dot{\Theta}|^4 - \sum_{i=1}^{\infty} \alpha_i^4 |\dot{\Theta}_i|^4 \right) \right] d\omega}{E_0^2}, \tag{4}$$

where $\dot{\Theta} = \mathcal{F}\{\dot{\theta}\}$, $\dot{\Theta}_i = \mathcal{F}\{\dot{\theta}_i\}$, $E_0^2 = \int_{-\infty}^{\infty} |\dot{\Theta}|^4 d\omega$ and $E_i^2 = \int_{-\infty}^{\infty} |\dot{\Theta}_i|^4 d\omega$.

The first term of Eq. (4) includes the contribution of the bandwidth of each harmonic to the bandwidth of the full signal, and the second term represents the contribution due to the interactions between the harmonics. It can be shown that the bandwidth of the full signal is mainly affected by that of the fundamental harmonic and the harmonic interaction term. Depending on the mechanical system, the type of nonlinearity and the coefficients $\alpha_i$, the computed nonlinear bandwidth can be more, less, or even equal to that of the baseline (approximately linear) system, as can be seen in Fig. 4c. Importantly, it should be noted that due to the fact that the harmonics and their intensities are energy-dependent, the contribution of each term in Eq. (4) varies with energy as well, highlighting the *energy tunability of the bandwidth of the nonlinear system* and the flow of energy in the frequency domain.

The precise mechanism is elucidated with the aid of Ext. Data Figs. 1 – 4. As discussed in the main text, in Case 2 the bandwidth of the system increases, but not by as much as the decay-time constant decreases, so the T-B product becomes less than unity. To demonstrate the significant decrease of the decay-time constant we consider the experimental response associated with an input energy of $\sim 10^{-2}$ J kg$^{-1}$ m$^{-2}$, and perform inverse wavelet transform harmonic decomposition [23] to separate the fundamental harmonic and the second harmonic (as a representation of all the higher harmonics). The combined response that includes the first and second harmonics and their corresponding wavelet spectra, are illustrated in Ext. Data Fig. 1. Comparing the amplitudes of the harmonics with the combined response, the figure indicates that a significant amount of the energy is carried by the second harmonic (and the rest of the higher harmonics). We note that the second and higher harmonics decay at a relatively short time, approximately 5s, which significantly shortens the decay-time constant (i.e., the storage time). This effect may be further verified by examining the instantaneous decay rates [23] of the fundamental and the second harmonic – cf. Ext. Data Fig. 2. These results show that, in addition to rapidly vanishing in ~ 5 seconds, the second harmonic has a much higher decay rate than the fundamental harmonic (which persists for more than 4 times longer compared to the second harmonic), which eventually results in a significant decrease of the overall decay-time constant of the response of the nonlinear system. We may arrive at the same conclusion by examining the decay-rate of the full signal and its storage time. In cases where the overall decay-rate (mean decay-rate) of the signal after the storage time (since the RMS bandwidth of a system is directly related to its overall decay-rate – in the case of linear systems, bandwidth and decay-rate are equal) is less (greater) than the reciprocal of its storage time ($\Delta\omega \lessgtr 1/\Delta t$), the time-bandwidth product of the system is less (greater) than unity. This interpretation indicates that for systems with time-bandwidth product above (below) unity, the decay-rate of the system increases (decreases) compared to that for time less than the storage time.

The storage time in the case of the signal in the Ext. Data Fig. 1a, the mean decay-rate (depicted in Ext. Data Fig. 2) of which is $0.24 \text{ s}^{-1}$ (where the bandwidth is 0.22), is $\Delta t = 2.4 \text{ s}$ (marked in Ext. Data Fig. 2). In this case we notice that the mean decay-rate, $0.24 \text{ s}^{-1}$ is less than the reciprocal of the storage time, $1/\Delta t = 0.42$, of the signal, resulting in time-bandwidth product of less than unity. This means that compared to while the energy of the system is reaching 37% of its maximum energy (storage time), the decay-rate of the system decreases.

Next, we turn our attention to Case 3, particularly to the situation corresponding to input energy of $\sim 10^{-2} \text{ J.kg}^{-1}.\text{m}^{-2}$ for which the T-B product becomes greater than unity. As stated earlier, in this case the decay-time constant of the system increases disproportionally compared to how its bandwidth decreases. Again, we may see why with the aid of the instantaneous decay rates of the harmonics. Ext. Data Fig. 3 shows the response of the system for the above input energy, its fundamental and second harmonics, and their corresponding wavelet spectra. By comparing the time series of the harmonics, one may observe that the fundamental harmonic is losing energy significantly slower than its counterpart in the Ext. Data Fig. 1, resulting in an increased decay-time constant. Notably, we see that the amplitude of the second harmonic in the beginning actually *increases*, which contributes to an increase in the storage time. Thus, contrary to Case 2, the 2nd harmonic (and the remaining higher harmonics), instead of more rapidly dissipating the energy of the system, actually stores the energy of the system for longer times. This is further verified by examining the decay rate of the fundamental and the second harmonics (Ext. Data Fig. 4), from which we observe that, unlike the decay rate of the fundamental harmonic that is always positive (meaning that it continuously loses energy), the decay rate of the second harmonic attains negative values at early times during which it actually gains energy; this implies that the second harmonic loses (transfer) energy to the fundamental harmonic. Overall, this nonlinear energy flow mechanism between the harmonics leads to a significant increase in the decay-time constant of the system in Case 3. Examining the mean decay-rate, $0.18 \text{ s}^{-1}$ (cf. Ext. Data Fig. 4), and the storage time, $\Delta t = 15.6 \text{ s}$ (marked in Ext. Data Fig. 4) for this case (cf. the full response in the Ext. Data Fig. 3a), we may see that the mean decay-rate is greater than the storage time reciprocal, implying a time-bandwidth product greater than unity. This means that compared to the time where the system is losing 63% of its energy (storge time), the decay-rate of the system increases significantly.

**A2. Experimental fixture assembly**

The experimental fixture consists of a pendulum made of a ¼-in diameter carbon fiber rod, 45-in long, which is bonded at one end to a 2-in diameter 306 stainless steel ball of Rockwell hardness C25. The pendulum is then rigidly attached to a ¼-in diameter (horizontal) stainless steel rod, 2.5-in long, which acts as a hinge. The hinge is then supported by a pair of SAE 841 Bronze flanged oil embedded sleeve bearings – cf. Ext. Data Fig. 5. For the experiments in which impacts with the rigid barrier (stop) occur, the pendulum ball impacts a 2×5×6-in³ steel block supported by a steel vise. The vise is placed so that the impact surface of the steel block is offset 1.18-in from the steel ball and is perpendicular to the plane of motion of the pendulum. Lastly, for the case where impacts occur in the presence of a magnetic field, i.e., Case 3, the fixture is modified by the addition of two pairs of magnets, 3 inches apart, placed next to the impact surface of the steel block, cf. Fig. 2, and another magnet attached to the bottom of the steel ball of the pendulum. The magnets are placed such that the pairs of magnets on the steel block repel the magnet on the pendulum with the goal of creating an unstable equilibrium point close to and in front of the steel block of the fixture, and, hence, introduce softening nonlinear effects in the dynamics of the pendulum.

Each test was performed as a free-response experiment with non-zero initial angle and zero initial angular velocity. To ensure such initial conditions, the pendulum was held at the desired initial angle by a 12VDC cylinder electromagnet and was released for each test by cutting the power to the electromagnet.

**A3. Data acquisition and postprocessing**

Due to the very low frequency of oscillation of the pendulum, i.e., ~0.5 Hz, typical accelerometers and vibrometers cannot be used to accurately measure its response. For this reason, we recorded (video-captured) the motion of the pendulum at 240 frames-per-second with full HD quality until it settled to its fixed point. For the purpose of tracking the motion of the pendulum, we covered the steel ball with white tape, placed a small black tracking point on it (cf. Fig. 2b & 2c), and filmed its motion at a 14-inch offset from the pendulum. After a video is recorded, we converted it to a series of gray-scale frames to speed up the data acquisition process. To decrease the noise from the capture, we placed a 60×60 pixel moving window on the tracking point to which we then applied a 2D smoothing Gaussian filter with standard deviation of 2. Once the motion of the pendulum is quantified, the coordinate system was placed at its fixed point and, considering the length of the pendulum, its pixel location was converted to an angle. To calculate the angular velocity, we differentiated the time series of the angle time series and suppressed the "numerical noise" caused by differentiation with a 3$^{rd}$ order lowpass Butterworth filter with cutoff frequency of 10 Hz. Finally, the angular acceleration was obtained by differentiating the filtered angular velocity.

Furthermore, to obtain the envelope of the angular velocity signal, we computed its absolute value and extracted its local maxima, which were at least 0.75 seconds apart and assumed values of at least 0.01 rad/s. Then, to obtain the envelope signal, the local maxima were interpolated by Akima Spline curves [24] to avoid extreme fluctuations and ensure C$^1$ continuity, and then were extrapolated to time $t = 0$ at the left and to the right until the envelope became zero. Once the envelope of the angular velocity signal was obtained, one could compute the bandwidth, decay-time constant, and time-bandwidth product of the system associated with the specific input that produced the angular velocity signal.

## A4. System identification and the experimental time-bandwidth plots

For Case 1 of the experiments, i.e., the case where no impacts occurred, two tests were performed, namely, one for system identification and another for validation. For the former, the initial angle was set to $\theta(0) = 6°$, and for the latter to $\theta(0) = 15°$. Due to presence of friction and the fact that the system was modeled as a single degree-of-freedom oscillator, we assumed the following reduced-order model for its oscillation,

$$\ddot{\theta} + \lambda\dot{\theta} + \omega_0^2\theta + \mu\,\text{sgn}(\dot{\theta}) = 0, \quad \theta(0) = \theta_0, \dot{\theta}(0) = 0, \tag{5}$$

where all coefficients were normalized with respect to the inertia ($I$) of the pendulum. By performing time series reconciliation, i.e., by matching the experimentally measured time series and the one predicted by the model (5), we obtained the normalized damping coefficient, $\lambda = 0.0056\ \text{s}^{-1}$, natural frequency, $\omega_0 = 2.8247\ \text{rad/s}$, and friction coefficient, $\mu = 0.0056\ \text{rad.s}^{-2}$ in (5). Using the identified values, we reproduced the response of the pendulum for the validation experiment with a coefficient of determination [25] of $R^2 = 0.9967$ – cf. Ext. Data. Fig. 6. The set of experiments for Case 1 not only allowed us to create an accurate computational model of the pendulum, but also provided the baseline time-bandwidth characteristics based on which we could highlight and assess clearly the effects of the inelastic impacts and the repelling magnetic forces on the said characteristics. The bandwidth and decay-time constants of this system are shown in Figs. 3a and 3b, with the time-bandwidth product being normalized by a factor of 1.75 to approach unity for small values of input energy. This normalization was performed to better depict and compare the time-bandwidth products of all experimental cases. It should also be noted that, for Cases 2 and 3, the corresponding time-bandwidth products were divided by the same normalization factor of 1.75 for fair comparisons with Case 1 and with each other.

Next, we considered Case 2 of our experiments, which is the oscillating pendulum with impact nonlinearity. As explained earlier, the impact nonlinearity is realized by the placing a steel block at a clearance of 3 cm away from the pendulum. This means that for low enough energies the oscillations of the pendulum could not overcame that clearance, resulting in behavior analogous to Case 1. For higher energies (larger initial angles) of the pendulum though, a finite number of inelastic impacts occurred. For this set of experiments, we considered model (5) and updated it to accommodate for the impacts. Each inelastic impact was modeled using a restitution coefficient less than unity, $r$, as,

$$-\frac{\dot{\theta}^+}{\dot{\theta}^-} = r < 1, \tag{6}$$

where superscripts in $\dot{\theta}^+$ and $\dot{\theta}^-$ denote the angular velocities before and after the impact, respectively. Considering (6), from our experiments we were able to compute the restitution coefficient to be $r \approx 0.6$. Therefore, using the identified $r$ and parameters from (5) we could computationally predict the time-bandwidth characteristics of the pendulum with (strong) impact nonlinearity – cf. Fig. 3.

Then, to experimentally validate and ensure the accuracy and reproducibility of the measurements and time-bandwidth tunability of the nonlinear system for Case 2, we performed a set of eight tests for each of the initial angles, $\theta_0 = 0.7°, 1.0°, 1.2°, 1.5°, 3.3°, 5.1°, 6.9°, 8.6°, 9.6°$ and $11.2°$. For each test we postprocessed the angular velocity of the system according to what we described before, and computed the input energy to the system, bandwidth, decay-time constant, and their normalized product. Finally, we computed the mean value and standard deviation of the eight tests per initial angle and superimposed them onto the time-bandwidth curves in Fig. 4. Similar to Case 2, in Case 3 we performed eight tests for each of the initial angles $\theta_0 = 1.9°, 2.3°, 2.5°, 2.9°, 3.3°, 4.6°, 6.2°, 6.9°, 8.6°, 9.6°$, and for each experimental test we computed the time-bandwidth characteristics of the pendulum with impact nonlinearity in the presence of a magnetic force.

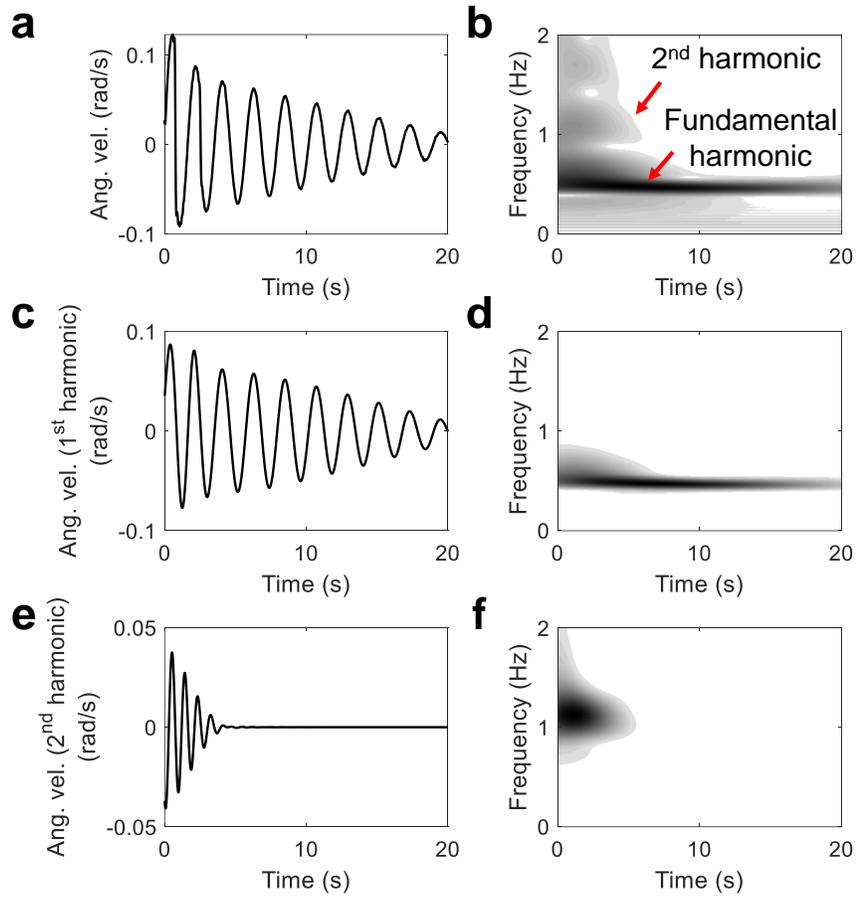

Extended Data Fig. 1. Harmonic separation – Case 2. Response of the experimental system excited by normalized input energy of $\sim 10^{-2}\,\mathrm{J\,kg^{-1}\,m^{-2}}$ in (a) the time domain and (b) with its corresponding wavelet spectrum. (c) and (d) show the time series and the wavelet spectrum of the fundamental harmonic, respectively. (e) and (f) show the time series and the wavelet spectrum of the second harmonic, respectively.

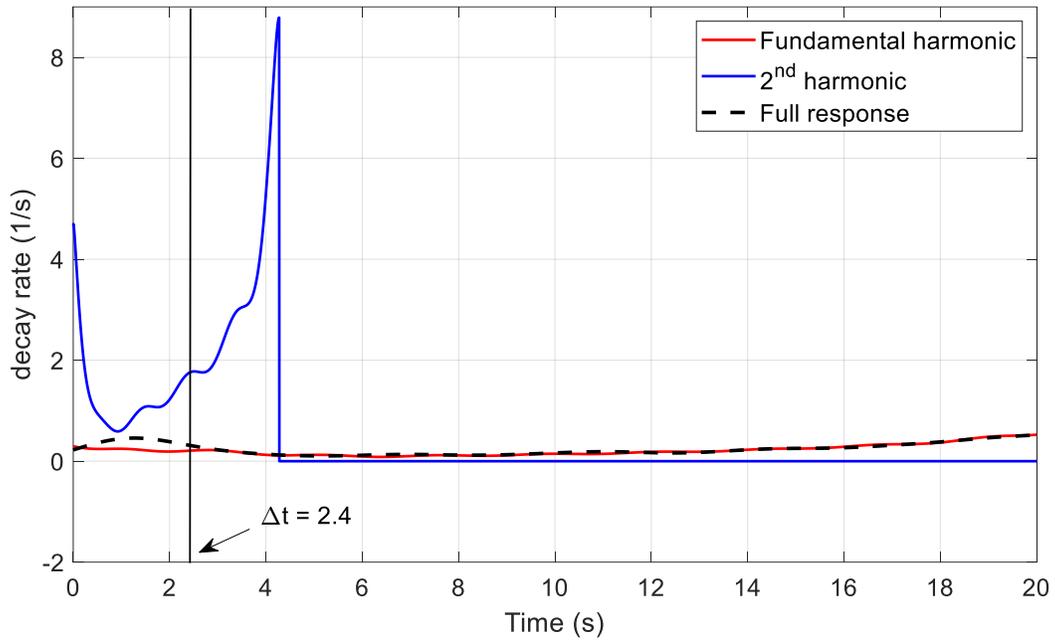

Extended Data Fig. 2. Decay rates of harmonics – Case 2. The black, the red and the blue curves represent the instantaneous decay rate of the full signal, fundamental and second harmonic, respectively, of the experimental measurements depicted in Ext. Data Fig. 1.

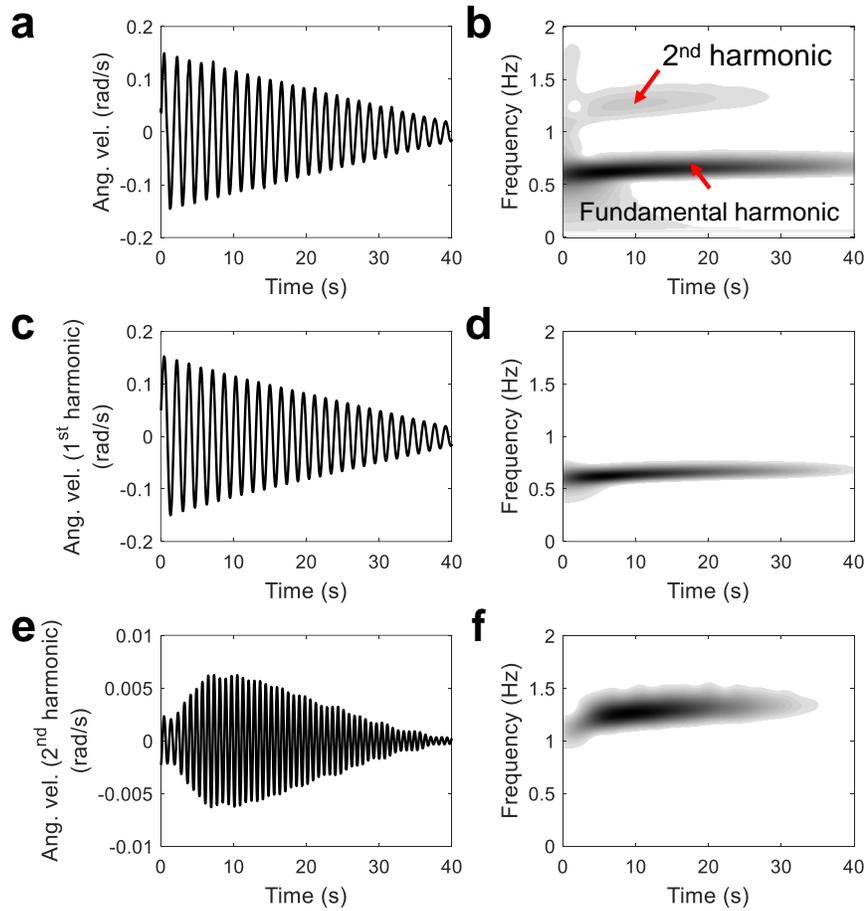

Extended Data Fig. 3. Harmonic separation – Case 3. Response of the experimental system excited by normalized input energy of $\sim 10^{-2}\,\mathrm{J\,kg^{-1}\,m^{-2}}$ in (a) the time domain and (b) with its corresponding wavelet spectrum. (c) and (d) shown the time series and the wavelet spectrum of the fundamental harmonic, respectively. (e) and (f) show the time series and the wavelet spectrum of the second harmonic, respectively.

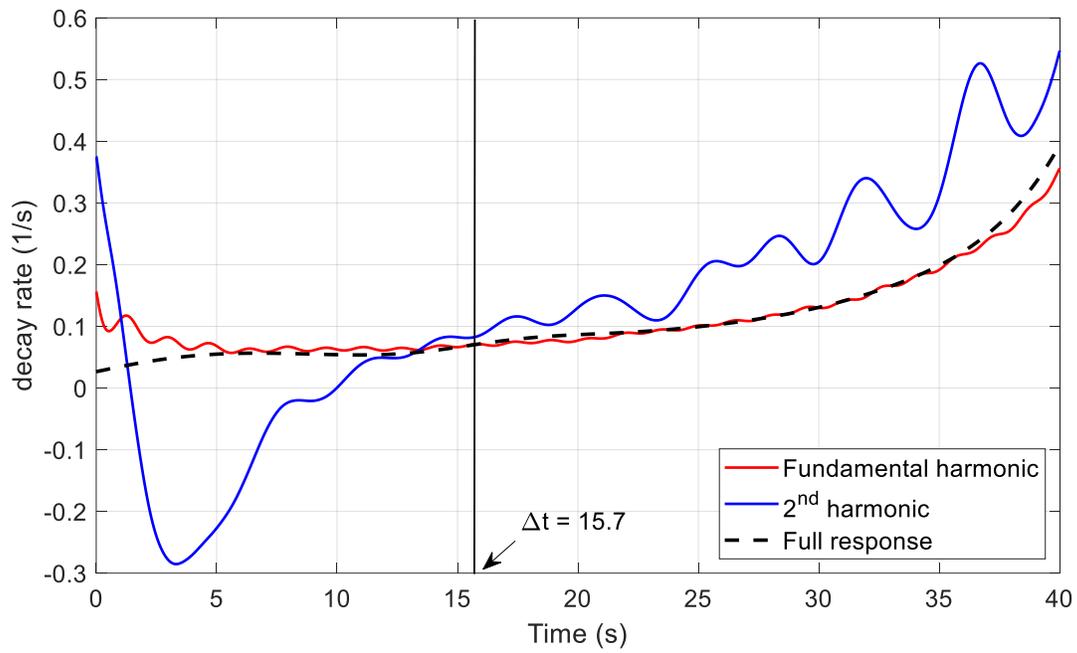

Extended Data Fig. 4. Decay rates of harmonics – Case 3. The black, the red and the blue curves represent the instantaneous decay rate of the full signal, fundamental and second harmonic, respectively, of the experimental measurements depicted in Ext. Data Fig. 3.

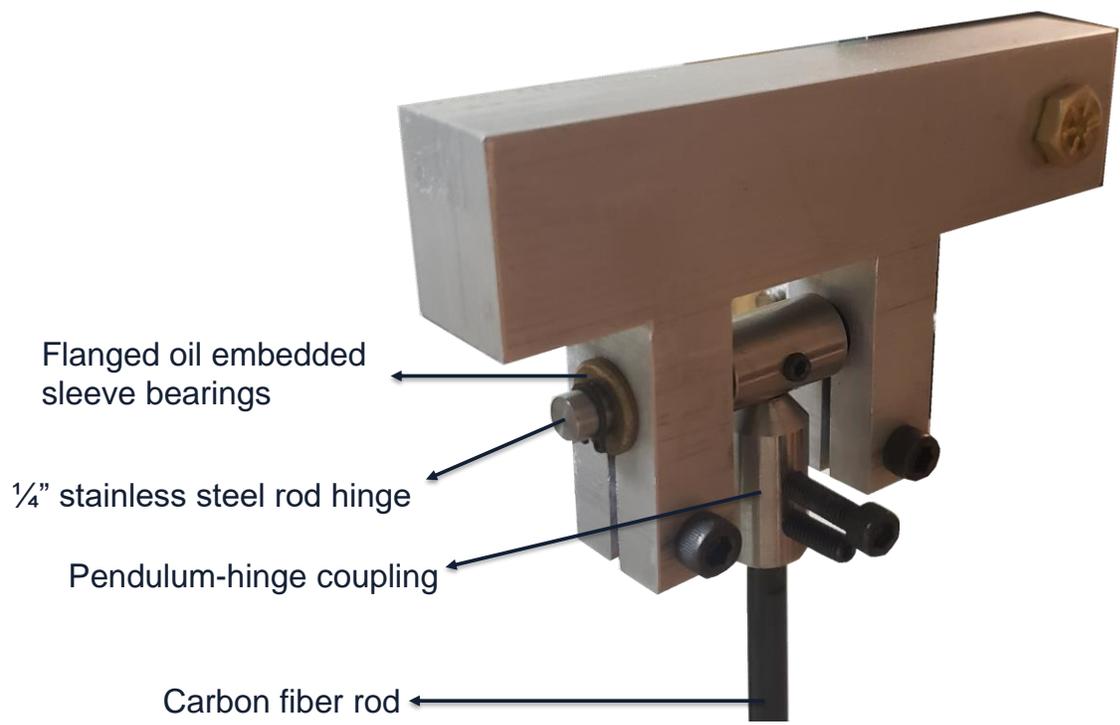

Extended Data Fig. 5. Configuration of the pendulum mount and pivot. Fully assembled mount of the pendulum, along with the hinge from which it hangs.

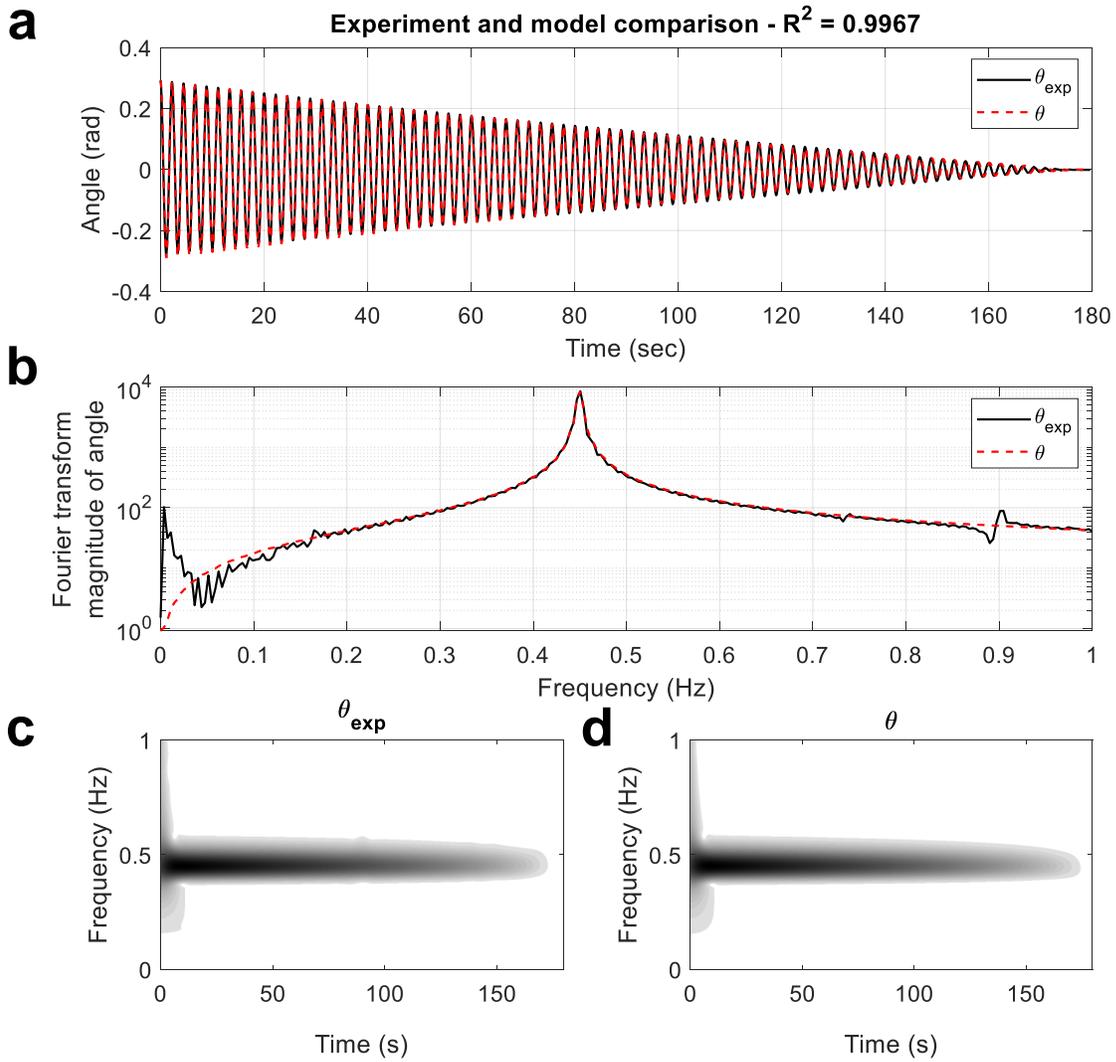

Extended Data Fig. 6. System identification validation. Comparison between (a) the computational reduced-order model (5) (red dashed curve) and the experimental response of the pendulum (black solid curve), showing significant accuracy, i.e., the coefficient of determination [25] is $R^2 = 0.9967$. (b) The corresponding Fourier transform of the time series depicted in (a). (c) and (d) illustrate the wavelet transform spectra of the response of the experimental response of the pendulum and the identified model (5), respectively.

# References

1. Tsakmakidis KL, Shen L, Schulz SA *et al.* (2017) Breaking Lorentz reciprocity to overcome the time-bandwidth limit in physics and engineering. *Science* **356**, 1260-1264.

2. Amoroso F (1980) The bandwidth of digital data signal. *IEEE Communications Magazine* **18**, 13-24.

3. Chaplain GJ, De Ponti JM, Aguzzi G *et al.* (2020) Topological Rainbow Trapping for Elastic Energy Harvesting in Graded Su-Schrieffer-Heeger Systems. *Physical Review Applied* **14**, 054035.

4. Fernandes DE, Silveirinha MG (2019) Topological Origin of Electromagnetic Energy Sinks. *Physical Review Applied* **12**, 014021.

5. Guglielmon J, Rechtsman MC (2019) Broadband Topological Slow Light through Higher Momentum-Space Winding. *Physical Review Letters* **122**, 153904.

6. Lu C, Wang C, Xiao M *et al.* (2021) Topological Rainbow Concentrator Based on Synthetic Dimension. *Physical Review Letters* **126**, 113902.

7. Mojahed A, Bunyan J, Tawfick S *et al.* (2019) Tunable Acoustic Nonreciprocity in Strongly Nonlinear Waveguides with Asymmetry. *Physical Review Applied* **12**, 034033.

8. Tsakmakidis KL, Boardman AD, Hess O (2007) 'Trapped rainbow' storage of light in metamaterials. *Nature* **450**, 397-401.

9. Tsakmakidis KL, Hess O (2012) Extreme control of light in metamaterials: Complete and loss-free stopping of light. *Physica B: Condensed Matter* **407**, 4066-4069.

10. Tsakmakidis KL, Hess O, Boyd RW *et al.* (2017) Ultraslow waves on the nanoscale. *Science* **358**, eaan5196.

11. Tsakmakidis KL, You Y, Stefański T *et al.* (2020) Nonreciprocal cavities and the time-bandwidth limit: comment. *Optica* **7**, 1097-1101.

12. Vakakis AF, Gendelman OV, Bergman LA *et al.* (2008) *Nonlinear Targeted Energy Transfer in Mechanical and Structural Systems*: Springer Netherlands.

13. Wang C, Kanj A, Mojahed A *et al.* (2020) Experimental Landau-Zener Tunneling for Wave Redirection in Nonlinear Waveguides. *Physical Review Applied* **14**, 034053.

14. Wang H, Wang Z, Li H *et al.* (2020) Signal evolution of an optical buffer based on the nonreciprocal silicon-on-insulator waveguide. *Optics Communications* **474**, 126158.

15. Xu J, Shen Q, Yuan K *et al.* (2020) Trapping and releasing bidirectional rainbow at terahertz frequencies. *Optics Communications* **473**, 125999.

16. Born M (1961) *Heisenberg und die Physik unserer Zeit*: Vieweg.

17. Gasparian V, Ortuño M, Schön G *et al.* (2000) Chapter 11 - Tunneling time in nanostructures. In *Handbook of Nanostructured Materials and Nanotechnology*, pp. 513-569 [HS Nalwa, editor]. Burlington: Academic Press.

18. Gabor D (1946) Theory of communication. Part 1: The analysis of information. *Journal of the Institution of Electrical Engineers-Part III: Radio and Communication Engineering* **93**, 429-441.

19. Hill M (2013) The uncertainty principle for Fourier transforms on the real line. *University of Chicago*.

20. Rao SS (2019) *Vibration of Continuous Systems*: Wiley.

21. Gardes FY, Tsakmakidis KL, Thomson D *et al.* (2007) Micrometer size polarisation independent depletion-type photonic modulator in Silicon On Insulator. *Opt Express* **15**, 5879-5884.

22. Sapsis TP, Dane Quinn D, Vakakis AF *et al.* (2012) Effective Stiffening and Damping Enhancement of Structures With Strongly Nonlinear Local Attachments. *Journal of Vibration and Acoustics* **134**.

23. Mojahed A, Bergman LA, Vakakis AF (2021) New inverse wavelet transform method with broad application in dynamics. *Mechanical Systems and Signal Processing* **156**, 107691.

24. Akima H (1970) A New Method of Interpolation and Smooth Curve Fitting Based on Local Procedures. *J ACM* **17**, 589–602.

25. Devore JL (2011) *Probability and Statistics for Engineering and the Sciences*: Cengage learning.